\documentclass[rnote]{aa}
\usepackage{graphicx}
\usepackage{txfonts}

\begin{document}

\title{On the Detectability of $^{57}$Fe Axion-Photon Mode Conversion in the Sun}

\author{J. Martin Laming\inst{1}}

\institute{Space Science Division, Naval Research Laboratory, Code 7674L, Washington DC 20375}

\abstract{}
{The purpose of this paper is to assess the feasibility of axion detection by X-ray spectroscopy of the sun.}
{We review the theory of axion-photon mode conversion with special attention to axions emitted
in the 14.4 keV M1 decay of $^{57}$Fe at the solar center. These then mode convert to photons in the outer
layers of the solar envelope, and may in principle be detected subsequently as X-rays.}
{For axion masses above about $10^{-4}$ eV, resonant mode conversion at a layer where the axion mass matches the local electron plasma frequency
is necessary. For axion masses above about $10^{-2}$ eV, this mode conversion occurs too deep in the solar atmosphere for the resulting photon to escape the solar surface and be detected before
Compton scattering obscures the line. At the (detectable) axion masses below this, the flux of mode converted
photons predicted by axion models appears to be too low for detection to be feasible with
current instrumentation. Nonresonant mode conversion for axion masses below
$10^{-4}$ eV is also plausible, but with still lower predicted fluxes, since the axion coupling constant is related to it mass.}
{Prospects for meaningful constraints on massive axion parameters from X-ray observations of this transition from the Sun do not appear to be promising. However parameters for massless counterparts (e.g. the ``arion'') may still result from such observations. It may mode convert in the outer layers of the solar atmosphere, but is not restricted by this to have a small coupling
constant.}{}

\maketitle

\section{Introduction}
The basic form of the Lagrangian for the theory of strong interactions known as
Quantum Chromodynamics (QCD) contains a pseudo-scalar term involving the gluon field strength
tensor that violates symmetry under the actions of charge and parity reversal (CP violation).
One practical consequence of this is the prediction of an electric dipole moment for the neutron in
excess of the experimental limits by about ten orders of magnitude.
The most generally accepted solution to this problem involves the introduction of a new global chiral symmetry (\cite{peccei77a,peccei77b}). \cite{weinberg78} and \cite{wilczek78}
pointed out that such a
symmetry, being spontaneously broken, necessitates the existence of a new boson, the axion, with mass in the range $10^{-12}$ eV to 1 MeV, depending on the scale at which the global symmetry is broken.
The axion mass is related to the symmetry breaking scale, $f_{PQ}$, by (e.g. \cite{moriyama95})
\begin{equation}
m=0.62\left(10^7 {\rm GeV}/f_{PQ}\right)\quad{\rm eV.}
\end{equation}

Limits on the axion mass are placed by various astrophysical constraints.
There are two most often discussed classes of axion; ``hadronic'' axions that do not couple to leptons and quarks
at tree level (\cite{kim79,shifman80}) and DFSZ axions that do (\cite{dine81,zhitnitskii80}).
Both types of axion are constrained to have masses less than about $10^{-3}-10^{-2}$ eV based on
analyses of the neutrino burst from SN 1987A (\cite{burrows89,keil97}), while hadronic axions
may also have a mass of a few eV, bounded from below by treatments of axion trapping in SN 1987A
(\cite{burrows90}) and from above by studies of red giant evolution (\cite{haxton91}).
Studies of white dwarf cooling and pulsation also indicate DFSZ axion masses of order $10^{-3}$ eV (\cite{isern10}). Cosmological arguments (e.g. \cite{kolb94}) give a lower limit on the axion
mass in the range $10^{-6} - 10^{-5}$ eV, with a mass of $\sim 10^{-5}$ eV supplying the measured
cold dark matter density (\cite{erken11}).

The sun is the only astrophysical object besides SN 1987A from which neutrinos have been directly
detected, and so it is natural to consider it also as a potential source of axions. While thermal
processes at the solar center produce axions by Compton
and Primakoff photoproduction processes, illustrated in Figure 1, (\cite{chanda88}), considerable attention has also been devoted to a transition at 14.4 keV in the $^{57}$Fe nucleus, which lies in a spectral region with essentially zero background from the nonflaring solar atmosphere. Being an M1 transition, it has a branching ratio for axion emission (DFSZ or hadronic)
instead of photon radiation (\cite{haxton91}), and has a sufficiently
low threshold to allow thermal excitation. An estimate of the axion flux in this line has been given by \cite{andriamonje09} as $F_A=4.56\times 10^{23}\left(g_{aN}^{eff}\right)^2$ cm$^{-2}$s$^{-1}$, where $g_{aN}^{eff}$ is the effective
axion-nucleon coupling. \cite{moriyama95} gives $F_A=5.3\times 10^8
\left(m/1 {\rm eV}\right)^2$ cm$^{-2}$s$^{-1}$ assuming the naive quark model (NQM) where $g_{aN}^{eff}=3.53\times 10^{-8} \left(m/1 {\rm eV}\right)$. \cite{andriamonje09} give $g_{aN}^{eff}=1.48\times 10^{-8}\left(m/1 {\rm eV}\right)$ (in the calculation of the $^{57}$Fe
M1 axion branching ratio), which would give a flux of
$F_A=10^8\left(m/1 {\rm eV}\right)^2$ cm$^{-2}$s$^{-1}$ but also discuss higher possible values for $g_{aN}^{eff}$.

\begin{figure}[h]
\centerline{\includegraphics[scale=0.65]{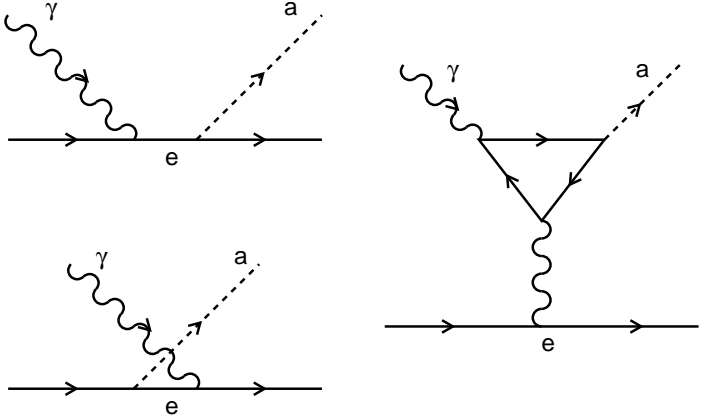}}
\caption{Feynman diagrams illustrating the two Compton scattering diagrams (left) and the Primakoff process (right).}
\end{figure}

\cite{andriamonje09} describe the CERN Axion Solar Telescope (CAST), designed to detect solar $^{57}$Fe axions by observing the 14.4 keV photons produced when the axions mode convert inside the detector. The parameters of CAST; a 9T magnetic field extending over a length of $s=9.26$ m, give a product $Bs\simeq 8\times 10^7$ G cm, which is smaller than the product of a typical solar magnetic
field and density scale height, and so the question of whether 14.4 keV photons might
be detectable in the solar spectrum as a signature of {\it in situ} axion mode conversion becomes
interesting to consider.

\section{Axion-Photon Interactions}
The axion-photon interaction Lagrangian is
\begin{equation}
L = -{g_{a\gamma\gamma}\over 4}aF^{\mu\nu}\tilde F_{\mu\nu} = g_{a\gamma\gamma}a{\bf E}\cdot {\bf B}
\end{equation}
where $g_{a\gamma\gamma}$ is the axion-photon coupling constant,
$a$ is the axion field, $F^{\mu\nu}$
is the electromagnetic field tensor and $\tilde F_{\mu\nu}$ is its dual. The product
$F^{\mu\nu}\tilde F_{\mu\nu}=-4{\bf E}\cdot {\bf B}$, which leads to the final form. Here, ${\bf E}$ is interpreted as the photon electric field and ${\bf B}$ the ambient magnetic field.
Thus axions only interact with photons when propagating across the ambient magnetic field, and
then only with the mode polarized along the magnetic field, known as the ordinary mode.

The effective Lagrangian for an axion-photon system is then
\begin{equation}
L = {1\over 2}{\bf E}\cdot {\bf \epsilon}\cdot {\bf E} -{1\over 2}{\bf B}\cdot {\bf B}
+{1\over 2}\left(\partial _{\mu}a\right)^2 -{1\over 2}m^2a^2 +g_{a\gamma\gamma}a{\bf E}\cdot {\bf B}
\end{equation}
where $m$  is the axion mass and
${\bf \epsilon} = 1-\omega _p^2/\omega ^2$ is the dielectric tensor in terms of the
electron plasma frequency $\omega _p$. We are working in the limit of small magnetic fields.
By way of contrast, \cite{yoshimura88} includes the magnetic field terms in ${\bf \epsilon}$.
The equations of motion are
\begin{eqnarray}
&{\bf \epsilon}\cdot {\bf E} - {k^2\over\omega ^2}{\bf E}+g_{a\gamma\gamma}a{\bf B}=0\cr
& {g_{a\gamma\gamma}{\bf E}\cdot {\bf B}\over \omega ^2}+\left(1-{m^2\over\omega ^2}-{k^2\over
\omega ^2}\right)a=0,
\end{eqnarray}
which for nontrivial solutions require
\begin{eqnarray}
n^2&={k^2\over\omega ^2}\cr
&={1\over 2}\left(\epsilon+1-{m^2\over\omega ^2}\right)\pm {1\over 2}
\sqrt{\left(\epsilon -1+{m^2\over\omega ^2}\right)^2+4{g_{a\gamma\gamma}^2B^2\over\omega ^2}}.
\end{eqnarray}
These two solutions represent axion and photon propagation across the magnetic field.
We rewrite the photon-axion mass matrix from equation 4 in symmetric form as
\begin{equation}
M = \left(\begin{array}{cc}
{m^2-\omega _p^2\over 2\omega}\quad & g_{a\gamma\gamma}B \\
g_{a\gamma\gamma}B\quad & -{m^2-\omega _p^2\over 2\omega}\\
                                             \end{array}
                                           \right).
\end{equation}
Pre and post multiplying by a rotation matrix,
\begin{eqnarray}
&\left(\begin{array}{cc}
\cos\theta\quad & -\sin\theta\\
\sin\theta\quad & \cos\theta\\ \end{array}\right) M
\left(\begin{array}{cc}
\cos\theta\quad & \sin\theta\\
-\sin\theta\quad & \cos\theta\\ \end{array}\right)=\cr
&\left(\begin{array}{cc}
{m^2-\omega _p^2\over 2\omega} \cos 2\theta -g_{a\gamma\gamma}B\sin 2\theta\quad &
{m^2-\omega _p^2\over 2\omega} \sin 2\theta +g_{a\gamma\gamma}B\cos 2\theta \\
{m^2-\omega _p^2\over 2\omega} \sin 2\theta +g_{a\gamma\gamma}B\cos 2\theta\quad &
-{m^2-\omega _p^2\over 2\omega} \cos 2\theta +g_{a\gamma\gamma}B\sin 2\theta \\
\end{array}\right),
\end{eqnarray}
which is diagonal for a mixing angle $\theta$ given by $\tan 2\theta = 2g_{a\gamma\gamma}B\omega/\left(\omega _p^2-m^2\right)$.

The evolution of the axion-photon mixed state with time $t$ is described by
\begin{eqnarray}
&\left|a\left(t\right)\right> = e^{-iE_at}\left|a^{\prime}\left(0\right)\right>\cos\theta +
e^{-iE_{\gamma}t}\left|\gamma ^{\prime}\left(0\right)\right>\sin\theta \cr
&=e^{-iE_at}\left|a\left(0\right)\right>\cos ^2\theta +
e^{-iE_{\gamma}t}\left|a\left(0\right)\right>\sin ^2\theta \cr
&+\cos\theta\sin\theta \left(e^{-iE_at}-e^{-iE_{\gamma}t}\right)\left|\gamma\left(0\right)\right>.
\end{eqnarray}
The probability of axion to photon conversion is then
\begin{eqnarray}
P\left(a\rightarrow\gamma\right)=\left|\left<\gamma\vert a\left(t\right)\right>\right|^2
&=\cos ^2\theta\sin ^2\theta\left|e^{-iE_at}-e^{-iE_{\gamma}t}\right|^2\cr
&=\sin ^22\theta\sin ^2{\left(E_a-E_{\gamma}\right)t\over 2}
\end{eqnarray}
which upon substituting for the mixing angle becomes
\begin{equation}
P\left(a\rightarrow\gamma\right)={g_{a\gamma\gamma}^2B^2\over g_{a\gamma\gamma}^2B^2+
\left(\omega _p^2-m^2\right)^2/4\omega^2}\sin ^2{\left(E_a-E_{\gamma}\right)t\over 2}.
\end{equation}
We put $\left(E_a-E_{\gamma}\right)t=\omega s\left(n_a - n_{\gamma}\right)$ with $n_{a,\gamma}^2$ given by the two solutions in equation 5, and $s$ the distance traveled through the medium. Then $E_a-E_{\gamma}\simeq\sqrt{\left(m^2-\omega _p^2\right)^2
/\omega ^2 +4g_{a\gamma\gamma}^2B^2}/2$ and
\begin{eqnarray}
P\left(a\rightarrow\gamma\right)&={g_{a\gamma\gamma}^2B^2\over g_{a\gamma\gamma}^2B^2+
\left(\omega _p^2-m^2\right)^2/4\omega^2}\times\cr
&\sin ^2{{s\over 2}\sqrt{\left(m^2-\omega _p^2\right)^2
/4\omega ^2 +g_{a\gamma\gamma}^2B^2}}\cr
&\rightarrow s^2g_{a\gamma\gamma}^2B^2/4\cr
\end{eqnarray}
when $\left(s/2\right)\sqrt{g_{a\gamma\gamma}^2B^2+\left(\omega _p^2-m^2\right)^2/4\omega^2}\rightarrow 0$.

The forgoing estimate assumed that the electron density in constant. When crossing a resonance, the rapid variation in $\tan 2\theta$ with $\omega _p^2$ first greater than $m^2$ becoming equal and then less than $m^2$, as the axion moves from high density to low density, means that the axion-photon oscillation cannot keep up with the changing Hamiltonian. In this limit, the equations of motion (4) admit solutions in terms of parabolic cylinder functions,
yielding an axion
transmission probability of $\exp\left(-\pi g_{a\gamma\gamma}^2B^2kh/\omega _p^2\right)$, (\cite{cairns83}; \cite{cally06}). The corresponding mode conversion probability is then
\begin{equation}
P\left(a\rightarrow\gamma\right)=1-\exp\left(-\pi g_{a\gamma\gamma}^2B^2kh/\omega _p^2\right)\simeq \pi g_{a\gamma\gamma}^2B^2kh/\omega _p^2,
\end{equation}
and as will be seen below, is generally larger than the nonresonant conversion probability given in equation 11.

\section{In Real Numbers ...}
We have used natural units, where $\hbar = c = 1$, above. To put variables in units more familiar to solar physicists, we use \cite{andriamonje09}
\begin{equation}
g_{a\gamma\gamma}=8.36\times 10^{-4}/f_{PQ}\quad {\rm GeV}^{-1},
\end{equation}
concentrating on DFSZ axions for the time being,
where $f_{PQ}$ is the symmetry breaking scale in GeV, and hence from equation 1
\begin{equation}
m=0.62\left(10^7 {\rm GeV}/f_{PQ}\right)
=0.74\left(g_{a\gamma\gamma}/10^{-10} {\rm GeV}^{-1}
\right) {\rm eV}.
\end{equation}
With these definitions
\begin{eqnarray}
g_{a\gamma\gamma}B&=10^{-16}B\left(g_{a\gamma\gamma}/10^{-10} {\rm GeV}^{-1}\right)\cr
&=1.3\times 10^{-16}B\left(m/1 {\rm eV}\right) {\rm cm}^{-1}
\end{eqnarray}
where $B$ is given in Gauss, and the conversion probability is
\begin{equation}
P\left(a\rightarrow\gamma\right)=4.5\times 10^{-33}B^2\left({\rm G}\right)\left(m/1 {\rm eV}\right)^2s^2.
\end{equation}
At a photon energy of 14.4 keV,
\begin{eqnarray}
\omega _p^2/\omega &= 4.8\times 10^{-21}n_e {\rm cm}^{-1}\cr
m^2/\omega &= 3.53 \left(m/1 {\rm eV}\right)^2 {\rm cm}^{-1}.\cr
\end{eqnarray}
For $m>10^{-4}$ eV and $s>10^8$ cm (a typical solar atmosphere scale size) the small angle approximation for the sine in equation
11 is not valid. Both terms in equation 17 are $>> g_{a\gamma\gamma}^2B^2$ and so unless $\omega _p\simeq m$, the conversion probability
is significantly reduced from the value in equation 16. The electron density
at which resonant mode conversion can occur is (see also \cite{yoshimura88})
\begin{equation}
n_e\simeq 7.35\times 10^{20} \left(m/1 {\rm eV}\right)^2.
\end{equation}
The electron scattering optical depth between this layer and the solar surface is
\begin{equation}
\tau \simeq n_e\sigma_T h\simeq 3\times 10^{-17}n_e
\end{equation}
where $\sigma _T$ is the Thomson cross section and $h\simeq 500$ km is the atmosphere density
scale height. Figure 2 shows emergent line profiles at the top of the solar atmosphere after
traveling through electron scattering optical depths of 0.1, 0.3, 1.0, 3.0, and 5.0, calculated
following the treatment of \cite{chandrasekhar60} and \cite{munch48}
for plane parallel atmospheres. As can
be seen, the detectable line disappears once $\tau\sim 5$, meaning a maximum density at which
detectable mode conversion will occur is $\sim 10^{17}$ cm$^{-3}$, with a maximum detectable
axion mass of $1.2\times 10^{-2}$ eV. At higher opacities, a broad Compton scattered feature
may still be seen. Figure 3 shows this feature at opacities of 5, 15, 50, and 150. It is visible
as a 1-2 keV wide feature out to an opacity of between 15 and 50. This would increase the maximum
detectable axion mass to about $4-7\times 10^{-2}$ eV, but would probably require more photons
detected due to the broader feature.

\begin{figure}[h]
\centerline{\includegraphics[scale=0.65]{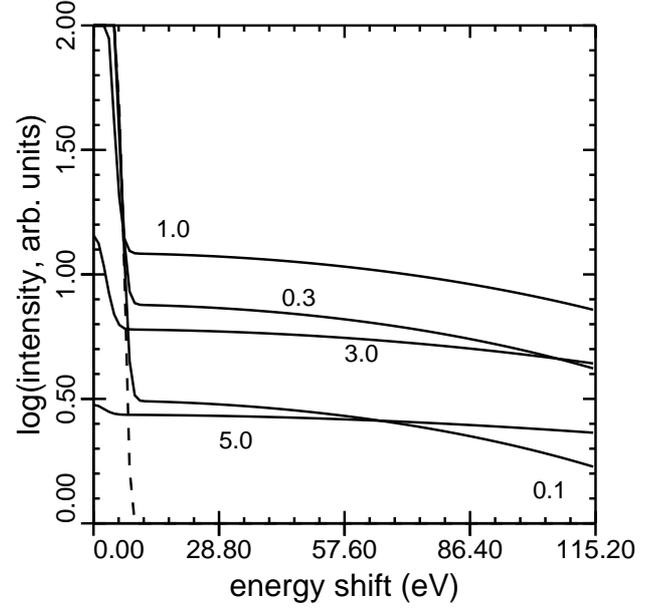}}
\caption{Line profiles of the emergent 14.4 keV $^{57}$Fe line, originally emitted as an axion,
and mode converted at various optical depths in the solar atmosphere. The electron scattering
opacities through which the various profiles have propagated are noted. The dashed curve gives
the original unscattered profile. At an opacity of $\simeq 5$, the central narrow line has
essentially disappeared.}
\end{figure}

\begin{figure}[h]
\centerline{\includegraphics[scale=0.65]{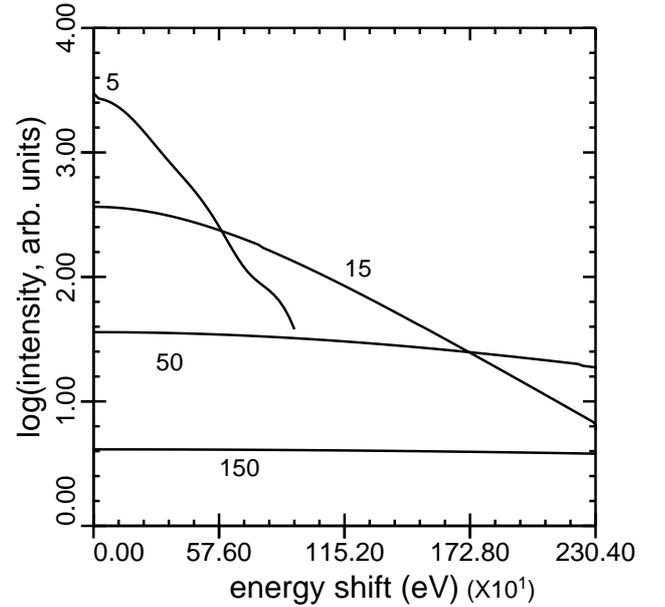}}
\caption{Same as Figure 1, but for electron scattering opacities of 5, 15, 50, and 150. The $\tau = 5$ curve is truncated for numerical reasons.}
\end{figure}

Equation 12 gives the mode conversion probability as $P = 7.5\times 10^{-25}B^2$,
independent of axion mass. For $B\sim 10^3$G, $P\sim 10^{-18}$, and the photon flux at
earth is $10^{-10}\left(m/{\rm 1 eV}\right)^2$ cm$^{-2}$s$^{-1}$, assuming the axion flux calculated by \cite{andriamonje09}. For axions with mass up to
$10^{-2}$ eV (the high end of the mass range allowed by SN 1987A observations, and close to the limit allowed by Compton scattering in the solar atmosphere), this leads to
a count rate of order $10^{-11}$ s$^{-1}$ in a instrument of effective area $10^3$ cm$^2$
(e.g. NUSTAR; \cite{harrison10,harrison13}); too low for detection. Significantly higher conversion probabilities would be
expected in pulsar strength magnetic fields, (\cite{yoshimura88}; \cite{perna12}).

For $s\sim 10^8$ cm, the argument of the sine in equation 11 becomes small for axion
masses at the low end of the allowed range, $m < 10^{-4}$ eV. In this case nonresonant conversion
can occur with probability $s^2g_{a\gamma\gamma}^2B^2/4\simeq 10^{-19}$ for $B\simeq 10^3$ G and $m\simeq 10^{-4}$ eV. This yields a flux at earth of mode converted photons of $5\times 10^4 \left(g_{aN}^{eff}\right)^2$. Taking $g_{aN}^{eff}=1.48\times 10^{-8}\left(m/1 {\rm eV}\right)$ as before gives a negligible mode converted photon flux at earth of $\sim 10^{-19}$ cm$^{-2}$s$^{-1}$ with $m=10^{-4}$ eV. The lower axion mass leads to a significantly lower flux, through its coupling to $g_{a\gamma\gamma}$.
In hadronic axion models, the parameters $g_{a\gamma\gamma}$ and $g_{aN}^{eff}$ are less tightly
constrained. \cite{yoshimura88} suggests $g_{a\gamma\gamma}$ may increase by a factor of ten,
which would lead to a commensurate increase in the count rates estimated above, though
still insufficient to change the conclusions above.

\begin{figure}[h]
\centerline{\includegraphics[scale=0.6]{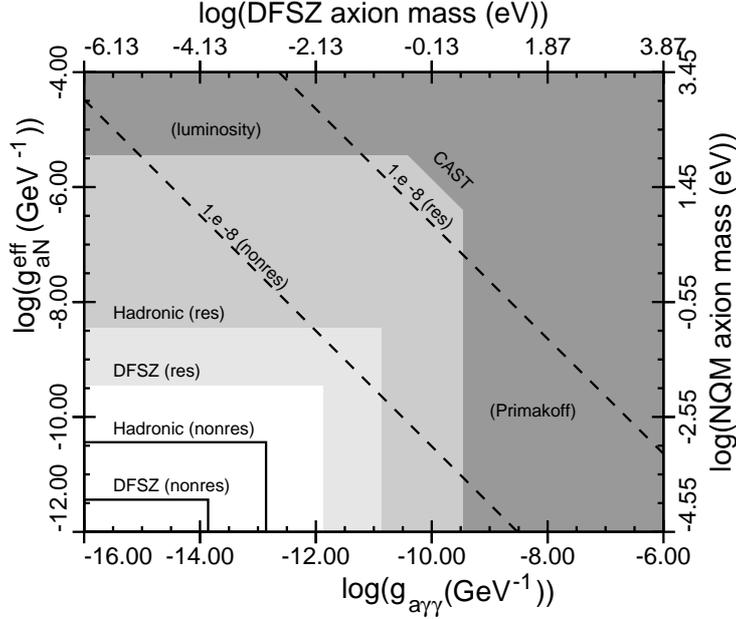}}
\caption{The $g_{a\gamma\gamma}-g_{aN}^{eff}$ parameter space. The dark region at the top and
right indicates the region ruled out by CAST (Andriamonje et al. 2009). The horizontal portion
labeled (luminosity) comes from a constraint on the total Fe$^{57}$ axion luminosity as a fraction
of the solar luminosity, and the vertical section labeled (Primakoff) comes from CAST constraints on axions produced by the Primakoff process, both from Andriamonje et al. (2009). The dashed contour below this indicates the where this limit moves with the nondetection of a count rate of $10^{-5}$ s$^{-1}$ by NUSTAR (a flux of $10^{-8}$ photons cm$^{-2}$s$^{-1}$) for resonant axion-photon
mode conversion. The shaded contours indicate the ``forbidden'' mass regions ($> 10^{-2}$ eV) towards the upper right for DFSZ or hadronic axions corresponding to the values of $g_{a\gamma\gamma}$ and $g_{aN}^{eff}$ (assuming the naive quark model). The lower dashed contour indicates the limit achieved by the same NUSTAR count rate for nonresonant axion-photon mode conversion in the outer layers of the solar atmosphere. The rectangular boxes indicate the ``allowed'' mass limits ($< 10^{-4}$ eV) for this case. These ``forbidden'' or ``allowed'' mass regions do not apply in the case of massless arions; only the CAST limits apply. Clearly, substantial parameters space is available to solar observations detecting nonresonant mode conversion of arions.}
\end{figure}

These potential constraints are illustrated in Fig. 4., showing the $g_{a\gamma\gamma}-g_{aN}^{eff}$ parameter space. The top and right hand axes show the corresponding axion masses in the cases of
DFSZ axions (equation 1), and those coming from the naive quark model from $g_{aN}^{eff}$ respectively. The dark region at the top and
right indicates the region ruled out by CAST (Andriamonje et al. 2009), and other constraints
discussed therein. The dashed contour below this indicates the where this limit moves with the nondetection of a count rate of $10^{-5}$ s$^{-1}$ by NUSTAR (a flux of $10^{-8}$ photons cm$^{-2}$s$^{-1}$, and about 10 times the background count rate in this rate estimated from Table 2
in Harrison et al. 2013) for resonant axion-photon
mode conversion. The shaded contours indicate the allowed mass region ($< 10^{-2}$ eV)in the lower left for DFSZ or hadronic axions. The lower dashed contour indicates the limit achieved by the same NUSTAR count rate for nonresonant axion-photon mode conversion in the outer layers of the solar atmosphere. The rectangular boxes indicate the mass limits ($< 10^{-4}$ eV) for this case. As can
be seen, in either resonant or nonresonant cases, the expected flux at the appropriate axion mass is well below plausible detection limits.

An exception might arise in the case of pseudo-Goldstone bosons arising from spontaneously
broken family symmetry, which are naturally massless (\cite{wilczek82}), but may otherwise
have similar properties to the axion (\cite{anselm88}). \cite{berezhiani90a}, \cite{berezhiani90b}, and \cite{berezhiani91} give
further discussion of the origin of such a particle, dubbed the ``arion'', in spontaneously
broken family symmetry. Being massless, it may be detected following mode conversion anywhere in the solar atmosphere where the electron
scattering opacity $\tau < 1$, which roughly corresponds to the nonresonant case in Figure 4.
In this case $g_{a\gamma\gamma}$ and $g_{aN}^{eff}$ are disconnected from the mass, and the
forbidden regions of Figure 4. for DFSZ and Hadronic axions with mass above $10^{-2}$ eV no longer apply (the CAST forbidden
region is still applicable). In such a case, NUSTAR might provide limits parameters competitive
with those derived by other methods. For DFSZ axions, studies of the Primakoff process in massive stars limit $g_{a\gamma\gamma} < 0.8\times 10^{-10}$ GeV$^{-1}$ (\cite{friedland12}), a limit
that would be accessible to solar observations. A similar limit comes from CAST observations of
Primakoff solar axions (\cite{andriamonje07}).

As mentioned above, axions may also be produced by thermal processes at the solar center (\cite{chanda88}). Hadronic axions may also be produced by the Primakoff process. Compton scattering and electron-nucleus bremsstrahlung can also produce DFSZ axions.
The Primakoff process alone produces around 10 - 100 times
more axions (\cite{andriamonje09}) than the $^{57}$Fe decay, while the other two process increase the axion flux by approximately $10^3$ (\cite{moriyama95}; \cite{derbin11}). The count rate so produced according to axion models would be $\sim 10^{-8}$ s$^{-1}$ at a mass of
$10^{-2}$ eV, still too low for observational feasibility, but possibly adequate to detect
the arion. However these
axions, and consequently the photons produced by their mode conversion, dominantly have energies
in the 1 - 3 keV range (\cite{derbin11}). At these energies, the dominant opacity is photoelectric absorption rather than Compton scattering, and the depth of solar atmosphere from which mode converted photons may be detected is smaller. In this spectral band,
the sun also emits a thermal bremsstrahlung spectrum, as well as
spectral lines from active regions and flares, in contrast to the 14.4 keV region, where no solar emission is expected outside of flare. The fundamental problem would be that if a definitive detection is not expected, to be of value, the observation needs to be designed so that a null result produces a further restriction of the observationally allowed parameter space.
\cite{hudson12} discuss progress and limitations in this approach. \cite{davoudiasl06}
suggest observing axion mode conversion in the magnetic field of the earth, in such a manner that the earth itself blocks out the thermal continuum from the sun. While this estimate
uses the observationally allowed axion parameter space rather than a specific model to
arrive at a count rate, it does have the advantage that a null result would be more readily
interpreted.

\section{Conclusions}
The detection of photons from axion mode conversion within the solar atmosphere is
practically limited to axion
masses at the higher end of the mass window, $\sim 10^{-5} - 10^{-2}$ eV, (bounded by cosmological arguments from below, and by stellar luminosities and SN 1987A from above). Detection of solar axions is also necessarily limited to masses below $10^{-2}$ eV due to the
Compton scattering expected between the mode conversion layer (deeper in the solar atmosphere
for higher mass axions) and the solar surface. This restriction means
that current instrumentation is unable to provide meaningful constraints on theoretical
axion models,
because axion coupling
constants scale as the axion mass, $m$, and so the mode converted signal scales as $m^4$, leading to an intrinsically weak signal for low mass axions. A possible exception to this might a massless
counterpart to the axion, the arion, where $g_{a\gamma\gamma}$ and $g_{aN}^{eff}$ are now decoupled
from the mass, and consequently the ``forbidden'' regions in Figure 4. do not apply.

\acknowledgements
This work has been supported by basic research funds of the Office of Naval Research.

\end{document}